\documentstyle[sprocl]{article}

\input{psfig}
\def\be{\begin{equation}}
\def\ee{\end{equation}}
\def\bea{\begin{eqnarray}}
\def\eea{\end{eqnarray}}
\begin{document}

\title{Cluster mass estimation from lens magnification}

\author{ Eelco van Kampen }

\address{Theoretical Astrophysics Center,\\
Juliane Maries Vej 30, DK-2100 Copenhagen {\O}, Denmark\\E-mail: eelco@tac.dk} 

%%%%%%%%%%%%%%%%%%%%%%%%%%%%%%%%%%%%%%%%%%%%%%%%%%%%%%%%%%%%%%
% You may repeat \author \address as often as necessary      %
%%%%%%%%%%%%%%%%%%%%%%%%%%%%%%%%%%%%%%%%%%%%%%%%%%%%%%%%%%%%%%

\maketitle\abstracts{
The mass of a cluster of galaxies can be estimated from its lens magnification, 
which can be determined from the variation in number counts of background
galaxies. In order to derive the mass one needs to make assumptions for the
lens shear, which is unknown from the variation in number counts alone.
Furthermore, one needs to go beyond the weak lensing (linear) approximation
as most of the observational data is concentrated in the central parts of
clusters, where the lensing is strong.
By studying the lensing properties of a complete catalogue of galaxy cluster
models, one can find reasonable approximations about the lens shear as a
function of the lens convergence. We show that using these approximations 
one can fairly well reconstruct the surface mass distribution from the
magnification alone.} 

\section{Motivation}

A rich cluster of galaxies acts as a gravitational lens on the galaxy
distribution beyond it. This simple fact can be used to derive a great
deal about both the lensing cluster as well as the background galaxy
population. Here we discuss how to best exploit the variation in galaxy
number counts caused by the lensing cluster, which enables one to derive
the lens magnification~\cite{btp}.
Recently, is has been shown that a depletion in number counts can clearly
be observed for the clusters Cl 0024+1654~\cite{fmd} and A1689~\cite{tetal}.

However, in order to obtain a mass for the lens, or
even a mass distribution, one needs the lens shear as well, because
the lens magnification $\mu$, shear $\gamma$ and convergence $\kappa$
(the dimensionless surface mass density) are related as
$\mu^{-1}= (1-\kappa)^2-\gamma^2$.

There are ways for obtaining the shear from observations~\cite{ks},
but one would like to obtain a mass
estimate that is independent of other methods. This means we need to find
assumptions for the shear, either as a function of surface mass density,
or as a function of the lens magnification.
We use a sample of numerical galaxy cluster models~\cite{vkk}
to find heuristic relations between $\gamma$ and $\kappa$ (or $\mu$),
some of which will have an underlying assumption
about the physical state of the lens, like isotropy.

\section{Estimating the lens convergence from lens magnification}

We wish to find a local relation for $\gamma$ vs.\ $\kappa$ and/or $\mu$.
The simplest one is the weak lensing approximation,
$\kappa_{\rm lin}=(1-\mu)/2$, which is valid only in the linear,
small-$\kappa$ regime~\cite{btp}, i.e.\ in the outskirts of clusters.
However, most observational data is available for the central
parts of clusters, where this approximation is invalid, and we need
to go beyond the weak lensing approximation to realistically estimate
the cluster surface mass density.

There are only two local relations between $\gamma$ and $\kappa$ that
result in a single caustic solution of the magnification equation
which is easily invertible~\cite{vk}: $\gamma=0$, corresponding
to a sheet of matter, and $\gamma=\kappa$, for an isotropic lens.
In the shearless case we have the estimate $\kappa_0=1-{\cal P}|\mu|^{-1/2}$,
while for the isotropic case the estimator becomes 
$\kappa_1=(1-{\cal P}|\mu|^{-1})/2$, where the ${\cal P}$ is the image
parity, i.e.\ the sign of $\mu$. Note that one can only measure $|\mu|$,
and therefore ${\cal P}$ has to be assigned by hand.
% Linearizing either one of these results in the linear approximation.

In practice, substructure and asphericity of the cluster will induce extra
shear~\cite{bsw}, especially in the
surrounding low-$\kappa$ neighbourhood, where substructure is relatively
more dominant and filaments make the cluster most aspherical. 
An approximation that tries to take these cluster lens features into
account, while still giving an invertible $\mu(\kappa)$ relation, is
$\gamma = \sqrt{(c+c^{-1}-2)\kappa}$.
This results in the amplification relation
$\mu^{-1} = (\kappa-c)(\kappa-c^{-1})$, with caustics at $\kappa=c$ and
$\kappa=c^{-1}$. The solution for $\kappa$ is then~\cite{vk}
\bea
\kappa_c = {c+c^{-1}\over2} + {\cal S}
     \Bigl[\Bigl({c+c^{-1}\over2}\Bigr)^2 - 
     {\cal P}|\mu|^{-1}-1\Bigr]^{1/2}\ ,
\label{eq1}
\eea
where we have introduced a second parity ${\cal S}$ which is the sign
of $\kappa-(c+c^{-1})/2$. Note that the $\gamma=0$ approximation is
recovered by setting $c=1$.

\begin{figure}
\psfig{figure=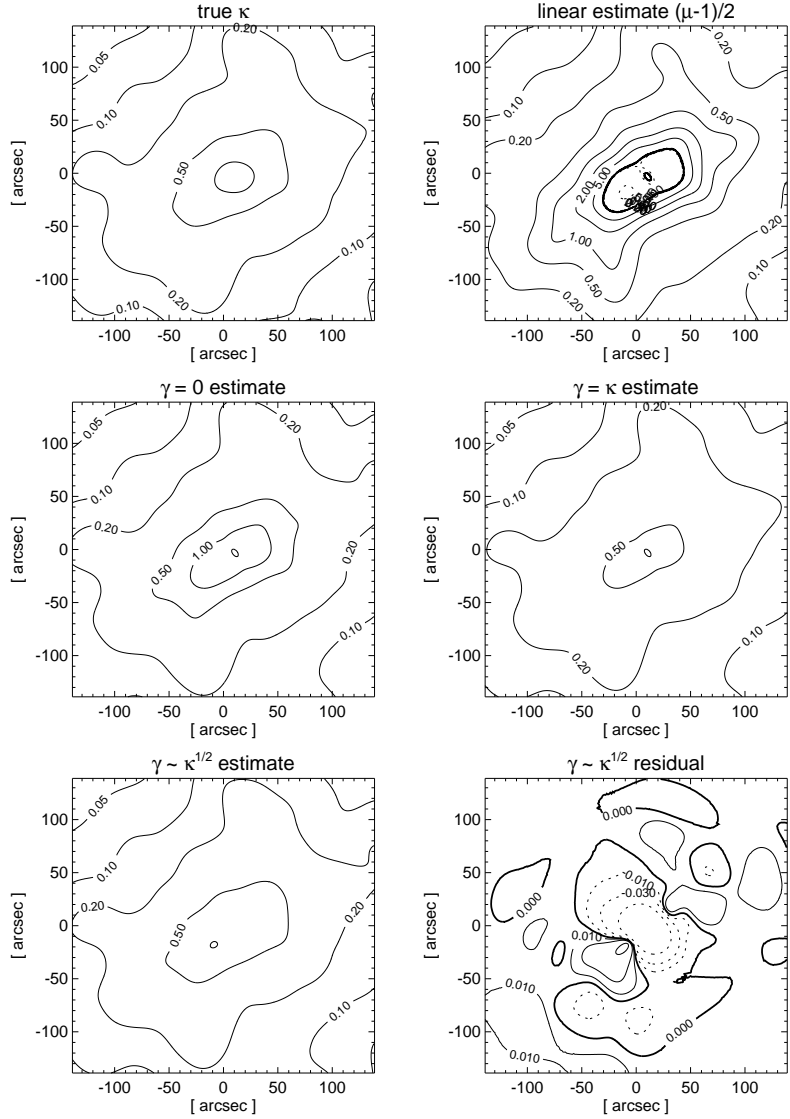,height=15cm}
\caption{True versus estimated $\kappa$-maps for a numerical model of
a rich cluster. The top-left panel shows the true $\kappa$, the next
four panels show the estimates for the linear, $\gamma=0$,
$\gamma=\kappa$, and $\gamma\propto{\kappa}^{1/2}$ approximations,
while the bottom-right panel shows the difference of the latter estimate
compared to the true distribution (i.e.\ top-left panel minus bottom-left
panel).}
\label{fig1}
\end{figure}

\section{Testing the convergence estimators on cluster models}

Although the feasibility of actually obtaining $\mu$ from real data is
an interesting topic for discussion, the issue here is how to
proceed from the measured magnification to other properties of the
lensing cluster, notably its mass distribution.
We therefore assume here that one can reliably measure the lens
magnification, and investigate how well the approximations allow us
to reconstruct the lens convergence from just this magnification map.
This should show us the best possible result each approximation can
provide us, and reveals systematics.

Using the thin lens approximation~\cite{bw},
we produce maps of the lens convergence, shear and magnification for
the cluster models. We then use the magnification only
(with full knowledge of its parity, though), to reconstruct convergence
maps using the various assumptions about the shear, and compare these
to the true convergence. This is shown in Fig.\ 1 for the most massive
model cluster from the catalogue, as this one has the largest range of
possible values for the convergence $\kappa$.

The linear estimator performs very poorly, as it
just follows the magnification, including the caustics. It is
only doing well for small $\kappa$, as expected~\cite{btp}. The $\gamma=0$
assumption produces an overestimate for the convergence for all regions
of the cluster. The $\kappa=\gamma$ estimator underestimates the mass
in the central regions of the cluster, and (slightly) overestimates
for $\kappa<0.2$. The $\gamma\propto{\sqrt\kappa}$ estimator clearly
performs best.

\section{Discussion}

The fact that these estimators can reproduce the surface mass density
reasonable well for model clusters does not guarantee that they work
on observational data. The best performing estimator has the clear
disadvantage that two parities need to be set.
More generally, one has to deal with many intrinsic sources of error
associated with observed magnification maps. Because the number of
background galaxies that can be used to construct a magnification map
will be finite, there will be shot noise.
Therefore one needs to smooth the distribution of number counts,
or average over annular bins. Furthermore, the backgrond galaxies
are clustered and have different redshifts, which produces extra
uncertainty. However, these problems can be dealt with to some
extent~\cite{btp,td}, and the magnification method, employing our
strong lensing estimators, seems a feasible way to directly measure
the dark matter distribution of galaxy clusters.

% \section*{Acknowledgments}

\end{document}